\documentclass [aps, pra, twocolumn, twoside, 10pt, amsfonts, amsmath, nofootinbib] {revtex4-1}

\usepackage {physics}
\usepackage{amsmath}
\usepackage{amsfonts}
\usepackage{graphicx}
\usepackage{float}

\newcommand{\Ising}{{\textrm{Ising}}}
\newcommand{\gauge}{{\textrm{gauge}}}
\newcommand{\vv}{{(v,v')}}
\newcommand{\vvv}{{(v,v'')}}

\newcommand{\config}{{\textrm{config.}}}

\begin {document}

\title {Dual $Z_2$ Lattice Gauge Theory of the 3D Ising Model with both Nearest- and Next-Nearest-Neighbor Couplings}

\author {Changnan \surname {Peng}} \email {Cppeng@Caltech.edu}
\affiliation {Department of Physics, California Institute of Technology, Pasadena,~California~91125,~USA}

\date{\today}

\begin {abstract}
  It is known that the normal three-dimensional (3D) Ising model on a cubic lattice is dual to the Wegner's 3D $Z_2$ lattice gauge theory. Here we find an unusual $Z_2$ lattice gauge theory which is dual to the 3D Ising model with not only nearest-neighbor (nn) coupling, but also next-nearest-neighbor (nnn) coupling. Our gauge theory has on each edge four $Z_2$ variables that have product $+1$, each located on a vector perpendicular to the edge. The nn coupling in the Ising model maps to the plaquette term in the gauge theory where the four variables multiplied have their vectors pointing inward, while the nnn coupling maps to the coupling between the $Z_2$ variables on nearby vectors on each edge in the gauge theory. A Wilson loop observable in the gauge theory depends on a framing of a loop, and maps to a surface of flipped-sign nn and nnn couplings in the Ising model. Further numerical simulations could be made to explore the universality at the phase transition.
\end {abstract}

\maketitle

\section {Introduction} \label {Introduction}

In high energy physics, symmetries play an essential role. The language used to describe the gauge symmetry is the gauge theory. Gauge theories accurately describe three of the four fundamental forces of nature - the electromagnetic force, the weak force, and the strong force~\cite{Gaillard1999}. However, some gauge theories such as quantum chromodynamics (QCD) are so complicated that it is very hard to solve them analytically. In order to solve these gauge theories, people discretize the spacetime into a lattice and use numerical methods to solve them. The lattice gauge theories give people more insights at the level of elementary particles. For example, the confinement of quarks can be displayed by the scaling property of Wilson loops in the lattice QCD~\cite{Wilson1974}. As a simpler lattice gauge theory, the $Z_2$ lattice gauge theory, which describes the symmetry when the field $\phi$ is changed to its negation $-\phi$, also has interesting properties with Wilson loops~\cite{Wegner1971,Wegner2014}.

	In 1971, F. J. Wegner discovered the duality of the $Z_2$ lattice gauge theory and the Ising model with nearest-neighbor coupling in a three dimensional cubic lattice~\cite{Wegner1971,Kramers1941,Wannier1945}. Also, a Wilson loop in the $Z_2$ lattice gauge theory is dual to a surface of frustrated links in the Ising model~\cite{Wegner1971,Bill2013}. Ising models are statistical models in which spins with value $\{\pm 1\}$ are placed on the lattice nodes and there are interactions between the spins. People are interested in Ising models because they display phase transitions and they are easy to simulate numerically~\cite{Onsager1944}. They are also good models to investigate the critical phenomena at the phase transition point.

	The critical phenomena are the universal behaviors of the models at the phase transition point. For example, at the phase transition point of the Ising models, the correlation function of the square of the spins (or the plaquette for $Z_2$ lattice gauge theory) decays with the distance between the spins in a power law, and the exponent of this power law should be independent to the microscopic structure of the model~\cite{Pelissetto2002}. We would like to test this universality for different Ising models and their dual gauge theories. The 3D Ising model with both nearest- and next-nearest-neighbor couplings will be a good model to work with. By changing the ratio between the couplings, we can get a continuous spectrum of models. However, the dual gauge theory of such Ising model was not reported.

In this paper, we show an unusual gauge theory that is dual to the 3D Ising model with both nearest- and next-nearest-neighbor couplings. In our gauge theory, each edge is associated with four $Z_2$ variables, each located on a perpendicular vector. The nearest- and next-nearest-neighbor couplings are dual to the four-vector plaquette coupling and the two-vector on-edge coupling, respectively. The duality between the defect bonds in the Ising model and the Wilson loops in the gauge theory is also discussed.



\section {Model} \label {Model}

The action of the 3D Ising model with both nearest-neighbor (nn) and next-nearest-neighbor (nnn) couplings is:
\begin{equation}
S_\Ising = -\beta_1\sum_{\vv\in nn}s_v^z s_{v'}^z - \beta_2\sum_{\vvv\in nnn}s_v^z s_{v''}^z, \label{S-Ising}
\end{equation}
where $v$, $v'$, and $v''$ are vertices in the 3D cubic lattice, $nn$ and $nnn$ respectively stand for the nearest- and next-nearest-neighbor vertices pairs, $s_v^z=\pm 1$ is the Ising variable placed on the vertex $v$, and $\beta_1$ and $\beta_2$ are the reduced coupling coefficients. In this paper, we only consider the cases when $\beta_1>0$ and $\beta_2\ge0$.

Before we talk about the action of the dual $Z_2$ lattice gauge theory, we first need to introduce the setup of the variables in this unusual lattice gauge theory.

The dual $Z_2$ lattice gauge theory has Ising variables with values $\pm 1$ placed on the edges of the dual cubic lattice. Each edge is associated with four variables, which are placed at the four sides of the edge. We show these variables as four perpendicular vectors to the edge, as shown in Fig.~1 and Fig.~2. (The vectors are only for illustrating the relative position of the variables. The values of the variables are not vectors; they are $\pm1$.)

There is a constraint on these variables: the product of four variables on each edge should equal to $+1$; i.e. there are even numbers of $+1$'s and $-1$'s among the four variables on each edge.

\begin{figure}
  \centering
  \includegraphics[width=7cm]{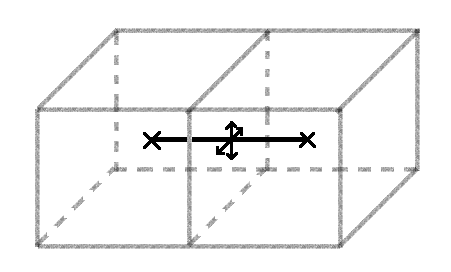}  
  \caption{The variables in the dual model. The grey cubes denote the lattice in the Ising model. The crosses at both ends of the black line denote the centers of the grey cubes. (The crosses are the vertices of the dual lattice.) The black line is the edge of the dual lattice. Four variables with values $\pm 1$ are placed on the black line, denoted as the four vectors in the middle of the black line.}
\end{figure}

\begin{figure}
  \centering
  \includegraphics[width=7cm]{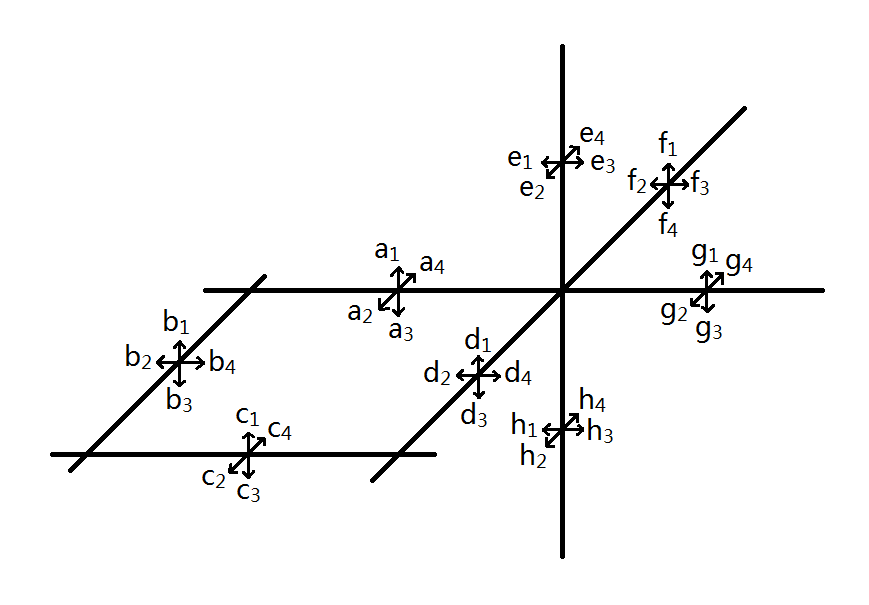}  
  \caption{More variables in the dual model. The black lines are the edges of the dual cubic lattice. The $a_i$'s, $b_i$'s, etc. are the variables in the dual model. Their values can be either $1$ or $-1$, and they need to satisfy the constraint - the product of four variables on each edge should be $1$, e.g. $a_1 a_2 a_3 a_4 = b_1 b_2 b_3 b_4 = \cdots = 1$.}
\end{figure}

The action of the dual gauge model is:
\begin{equation} \label{S-gauge}
S_\gauge = - \alpha_1\sum_{\textrm{plaquettes}}[a_2 b_4 c_4 d_2] - \alpha_2\sum_{\textrm{edges}}([a_1 a_2] + [a_2 a_3]),
\end{equation}
where ``$[\cdot]$'' means an example of the general idea, the ``$[a_2 b_4 c_4 d_2]$'' in the first term represents the product of the four variables on the edges of a plaquette, each pointing inward. (See Fig.~2 for the meaning of the letters.) The ``$[a_1 a_2]$'' or ``$[a_2 a_3]$'' in the second term represent the product of two nearby (but not opposite) variables on one edge. Because of the constraint on the four variables on an edge, i.e. $a_1 a_2 a_3 a_4 = 1$, there are $a_1 a_2 = a_3 a_4$ and $a_2 a_3 = a_4 a_1$. Therefore, the second term is actually symmetric: $a_1 a_2 + a_2 a_3 = (a_1 a_2 + a_2 a_3 + a_3 a_4 + a_4 a_1)/2$.

\begin{figure}
  \centering
  \includegraphics[height=2cm]{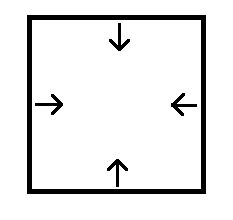}
  \includegraphics[height=2cm]{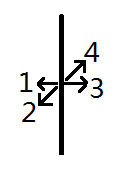}
  \caption{The first term in Eq.~\ref{S-gauge} is shown by the plaquette on the left. The four variables on the edges of the plaquette should all point inward. The order of the subscripts in the second term in Eq.~\ref{S-gauge} is shown on the right.}
\end{figure}

The first and second terms in the dual action (Eq.~\ref{S-gauge}) are better illustrated in Fig.~3. Some examples of the second term are shown in Fig.~4.

The coupling coefficients satisfy the dual relations:
\begin{eqnarray}
\sinh(2 \beta_1)\sinh(2 \alpha_1) &=& 1, \\
\sinh(2 \beta_2)\sinh(2 \alpha_2) &=& 1.
\end{eqnarray}

Note that when $\beta_2 = 0$, i.e. there is no nnn coupling, we will get $\alpha_2 = \infty$, i.e. the second term in the dual action dominates and makes the system always at its ``most probable state'' (the state with the minimal action). From the examples of the second term in Fig.~4 we can see that the ``most probable states'' are when the four variables on each edge have the same value, which means we can assign a single value to each edge. This goes back to Wegner's $Z_2$ lattice gauge theory~\cite{Wegner1971}.

\begin{figure}
  \centering
  \includegraphics[width=5cm]{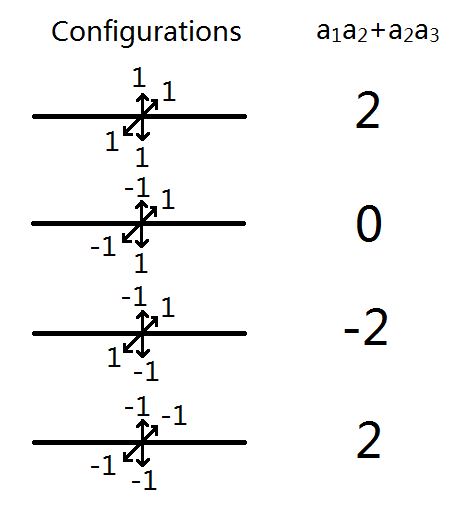}
  \caption{The possible configurations of the four variables on an edge, and their corresponding contributions in Eq.~\ref{S-gauge}. Note that the ``most probable states'' are when $a_1 a_2 + a_2 a_3$ reaches the maximum value, i.e. the four variables being the same.}
\end{figure}

In our unusual gauge theory, the local gauge transformation is to flip the values of the variables on the edges that all connect to one vertex in the dual lattice. As in Fig.~2, it means that we change the signs of the $a_i$'s, $d_i$'s, $e_i$'s, $f_i$'s, $g_i$'s, and $h_i$'s. It is not hard to check that Eq.~\ref{S-gauge} is invariant under this local gauge transformation.

The first and second terms in Eq.~\ref{S-gauge} are basic gauge invariants in our gauge theory. We can multiply the basic gauge invariants to get more complicated gauge invariants. Since the square of an Ising variable is always $1$, we can ``glue'' the plaquette terms with the edge terms in Eq.~\ref{S-gauge}, and erase the vectors that appear twice. Repeating this process, we can see that any loop of edges with one vector on each edge is gauge invariant. Some examples are shown in Fig.~5.

\begin{figure}
  \centering
  \includegraphics[width=7cm]{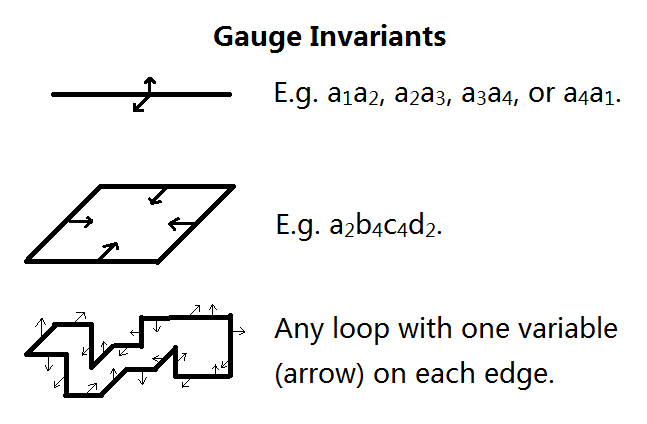}
  \caption{Examples of gauge invariants. We can decompose a single-vector-framing loop into basic gauge invariant terms by tilling the loop with plaquette terms and then changing the direction of the vectors on the framing by multiplying edge terms.}
\end{figure}

\section {Proof of Duality} \label {Proof}

To prove the duality, we consider the high temperature expansion (HTE) of one model and the low temperature expansion (LTE) of the other.

The HTE partition function of the Ising model is:
\begin{eqnarray} \label{Z-Ising}
Z_\Ising &=& \sum_\config e^{-S_\Ising}   \nonumber \\
&=& \sum_\config \exp\left(\beta_1\sum_{nn}s_v^z s_{v'}^z + \beta_2\sum_{nnn}s_v^z s_{v''}^z\right)   \nonumber \\
&\propto& \sum_\config \prod_{nn}\left(1+s_v^z s_{v'}^z\tanh\beta_1\right)   \nonumber \\
&& \qquad \times\prod_{nnn}\left(1+s_v^z s_{v''}^z\tanh\beta_2\right).
\end{eqnarray}
In the last step, we have use the trick that $\exp(\beta s)=\cosh\beta\ (1 + s\tanh\beta)$ when the value of $s$ can only be $\pm1$.

Because there is a sum over all configurations of the $s_v^z$'s in Eq.~\ref{Z-Ising}, after the expansion of the product, any term that does not cancel out all $s_v^z$'s will be eliminated by the sum.

The terms that cancel out all $s_v^z$'s are either $1$, or the loops constructed by the nn and nnn bonds. The basic element of such loops is the triangle with two nn bonds and one nnn bond. Such a basic loop contributes a term of $\tanh(\beta_1)^2 \tanh(\beta_2)$ in the HTE of the Ising partition function. All larger HTE loops can be tilled by the basic triangle loops.

Therefore, the HTE Ising partition function becomes:
\begin{equation}
Z_\Ising \propto 1+\sum_{\textrm{basic triangles}}\tanh(\beta_1)^2 \tanh(\beta_2)+\cdots.
\end{equation}

The LTE partition function of the gauge theory is:
\begin{eqnarray} \label{Z-gauge}
Z_\gauge &=& \sum_\config e^{-S_\gauge}   \nonumber \\
&=& \left.e^{-S_\gauge}\right|_{\max} + \sum \left.e^{-S_\gauge}\right|_{\textrm{excitations}}   \nonumber \\
&\propto& 1+\sum \exp(S_\gauge|_{\min}-S_\gauge|_{\textrm{excitations}}).
\end{eqnarray}

The simplest excitation of our dual gauge theory is to flip the signs of a pair of nearby variables on an edge. (Not to flip just one variable, because there is a constraint.) This flipping of signs affects two plaquettes and one edge, and thus contributes a term of $\exp(-4\alpha_1-2\alpha_2)$ in the LTE the partition function.

Therefore, the LTE gauge partition function becomes:
\begin{equation}
Z_\gauge \propto 1+\sum_{\textrm{simplest excitations}}\exp(-4\alpha_1-2\alpha_2)+\cdots.
\end{equation}

There is a one-to-one correspondence between the basic triangle loops in the Ising model and the simplest excitations in the gauge theory. As shown in Fig.~6, the pair of flipped variables are on the edge that penetrates the triangle. The two variables are the ones that point towards the nn bonds. It is not hard to see that there are also correspondences between larger loops in the Ising model and higher excitations in the dual gauge theory.

Therefore, we can see that the two models are dual if we choose $\tanh(\beta_1)=e^{-2\alpha_1}$ and $\tanh(\beta_2)=e^{-2\alpha_2}$. These conditions are equivalent to the dual relations we have shown in the previous section (Eq.~3 and 4).

\begin{figure}
  \centering
  \includegraphics[width=7cm]{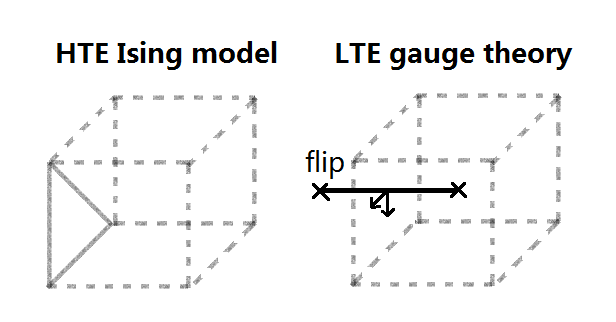}
  \caption{The duality of high temperature expansion (HTE) of the Ising model and the low temperature expansion (LTE) of the gauge theory.}
\end{figure}

Similarly, in the HTE of the gauge theory, we consider the ``loops'' (or ``wraps'') constructed by the terms in the dual action, i.e. the plaquettes and the two-vector edges. The unit of such a wrap is a cube in the dual lattice with six plaquettes and twelve two-vector edges. Such a basic wrap contributes a term of $\tanh(\alpha_1)^6 \tanh(\alpha_2)^{12}$ in the HTE gauge partition function.

As shown in Fig.~7, The corresponding LTE of the Ising model is to flip the sign of the spin that is at the center of the cube (or more generally, inside the wrap). This flipping of sign affects six nn bonds and twelve nnn bonds, and thus adds a term of $\exp(-12\beta_1-24\beta_2)$ in the LTE of the Ising partition function. There are also correspondences between larger wraps in the gauge theory and higher excitations in the Ising model. We can see that the two models are dual if we choose $\tanh(\alpha_1)=e^{-2\beta_1}$ and $\tanh(\alpha_2)=e^{-2\beta_2}$. These conditions are also equivalent to the dual relations Eq.~3 and 4.

\begin{figure}
  \centering
  \includegraphics[width=7cm]{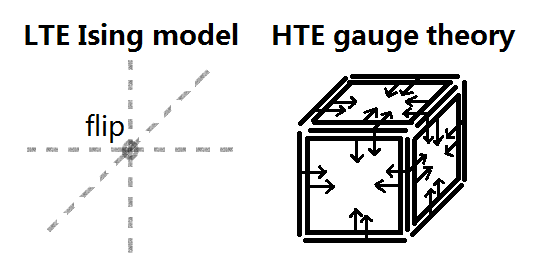}
  \caption{The duality of low temperature expansion (LTE) of the Ising model and the high temperature expansion (HTE) of the gauge theory. For clearness, not all plaquettes and edges are shown in the HTE of the gauge theory.}
\end{figure}

\section {Wilson Loop}

Like the Wilson loop in Wegner's $Z_2$ lattice gauge theory~\cite{Wegner1971,Wilson1974}, the gauge invariants in our gauge theory are also loops, but with vectors on the framing.

We start with the two basic gauge invariants in our gauge theory, the all-inward-vector plaquettes and the two-vector edges.

First, for a plaquette $a_2 b_4 c_4 d_2$ (see Fig.~2), the average value is:
\begin{eqnarray}
&&\frac{1}{Z_\gauge}\sum_\config a_2 b_4 c_4 d_2   \nonumber \\
&&\quad\times\exp\left(\alpha_1\sum_{\textrm{p}}[a_2 b_4 c_4 d_2] + \alpha_2\sum_{\textrm{e}}([a_1 a_2] + [a_2 a_3])\right)  \nonumber\\
&=& \sum_\config a_2 b_4 c_4 d_2 \prod_{\textrm{p}}(1+[a_2 b_4 c_4 d_2]\tanh\alpha_1)  \nonumber \\ &&\quad\times\prod_{\textrm{e}}(1+[a_1 a_2]\tanh\alpha_2)(1+[a_2 a_3]\tanh\alpha_2)  \nonumber \\
&&/\left\{\sum_\config \prod_{\textrm{p}}(1+[a_2 b_4 c_4 d_2]\tanh\alpha_1)\right.  \nonumber \\ &&\quad\left.\times\prod_{\textrm{e}}(1+[a_1 a_2]\tanh\alpha_2)(1+[a_2 a_3]\tanh\alpha_2)\right\}.
\end{eqnarray}

The expansion of the numerator can be split into two terms:
\begin{eqnarray}
&&\sum_\config a_2 b_4 c_4 d_2 \prod_{\textrm{p}}(1+[a_2 b_4 c_4 d_2]\tanh\alpha_1)  \nonumber \\ &&\quad\times\prod_{\textrm{e}}(1+[a_1 a_2]\tanh\alpha_2)(1+[a_2 a_3]\tanh\alpha_2)  \nonumber \\
&=&\sum_\config a_2 b_4 c_4 d_2 \ (1+a_2 b_4 c_4 d_2\tanh\alpha_1)  \nonumber \\
&&\quad\times\prod_{\textrm{p except } abcd}(1+[a_2 b_4 c_4 d_2]\tanh\alpha_1)  \nonumber \\ &&\quad\times\prod_{\textrm{e}}(1+[a_1 a_2]\tanh\alpha_2)(1+[a_2 a_3]\tanh\alpha_2)  \nonumber \\
&=&\sum_\config a_2 b_4 c_4 d_2 \prod_{\textrm{p except } abcd}(1+[a_2 b_4 c_4 d_2]\tanh\alpha_1)  \nonumber \\ &&\quad\times\prod_{\textrm{e}}(1+[a_1 a_2]\tanh\alpha_2)(1+[a_2 a_3]\tanh\alpha_2)  \nonumber \\
&&+\sum_\config \tanh\alpha_1 \prod_{\textrm{p except } abcd}(1+[a_2 b_4 c_4 d_2]\tanh\alpha_1)  \nonumber \\ &&\quad\times\prod_{\textrm{e}}(1+[a_1 a_2]\tanh\alpha_2)(1+[a_2 a_3]\tanh\alpha_2)  \nonumber \\
&=&\frac{1}{\tanh\alpha_1}\sum_\config a_2 b_4 c_4 d_2 \tanh\alpha_1  \nonumber \\
&&\quad\times\prod_{\textrm{p except } abcd}(1+[a_2 b_4 c_4 d_2]\tanh\alpha_1)  \nonumber \\ &&\quad\times\prod_{\textrm{e}}(1+[a_1 a_2]\tanh\alpha_2)(1+[a_2 a_3]\tanh\alpha_2)  \nonumber \\
&&+\tanh\alpha_1\sum_\config \prod_{\textrm{p except } abcd}(1+[a_2 b_4 c_4 d_2]\tanh\alpha_1)  \nonumber \\ &&\quad\times\prod_{\textrm{e}}(1+[a_1 a_2]\tanh\alpha_2)(1+[a_2 a_3]\tanh\alpha_2)  \nonumber \\
&=&\frac{1}{\tanh\alpha_1}\sum_{\textrm{wraps with }abcd}(\tanh\alpha_1)^{\cdots}(\tanh\alpha_2)^{\cdots}  \nonumber \\
&&+\tanh\alpha_1\sum_{\textrm{wraps without }abcd}(\tanh\alpha_1)^{\cdots}(\tanh\alpha_2)^{\cdots}.  \nonumber \\
\end{eqnarray}

There are two types of terms in Eq.~10: one corresponding to a wrap that contains the plaquette $a_2 b_4 c_4 d_2$, and the other does not. Note that a wrap in the gauge theory maps to the flipping of the Ising variables inside the wrap. We denote the two Ising variables which are closest to the plaquette $a_2 b_4 c_4 d_2$ as $v$ and $v'$, and note that they are on the two sides of the plaquette and $\vv\in nn$. In the first type of terms, where the corresponding wrap contains the plaquette, the two Ising variables on the two sides of this plaquette must be one inside the wrap and one outside the wrap, and thus they have different signs. On the other hand, in the second type of terms, where the wrap does not contain the plaquette, these two Ising variables must have the same sign. 

Using the dual relation Eq.~3, we can rewrite Eq.~10 in the Ising model side as
\begin{eqnarray}
&&e^{2\beta_1}\sum_{v,v'\textrm{diff. signs}}\exp(S_\Ising|_{\min}-S_\Ising) \nonumber \\
&&+e^{-2\beta_1}\sum_{v,v'\textrm{same sign}}\exp(S_\Ising|_{\min}-S_\Ising)  \nonumber \\
&=&\sum_\config e^{-2\beta_1 s_v^z s_{v'}^z}\exp(S_\Ising|_{\min}-S_\Ising).
\end{eqnarray}

Note that the denominator in Eq.~9 can be rewrite as
\begin{equation}
\sum_\config \exp(S_\Ising|_{\min}-S_\Ising).
\end{equation}
Therefore, Eq.~9, the average value of a plaquette, becomes
\begin{equation}
\frac{1}{Z_\Ising}\sum_\config e^{-2\beta_1 s_v^z s_{v'}^z}\exp(-S_\Ising),
\end{equation}
which is the average value of $e^{-2\beta_1 s_v^z s_{v'}^z}$. $v$ and $v'$ are the two vertices on the two sides of the plaquette. The values on the Ising model side are easier to compute in a numerical simulation.

\begin{figure}
  \centering
  \includegraphics[width=7cm]{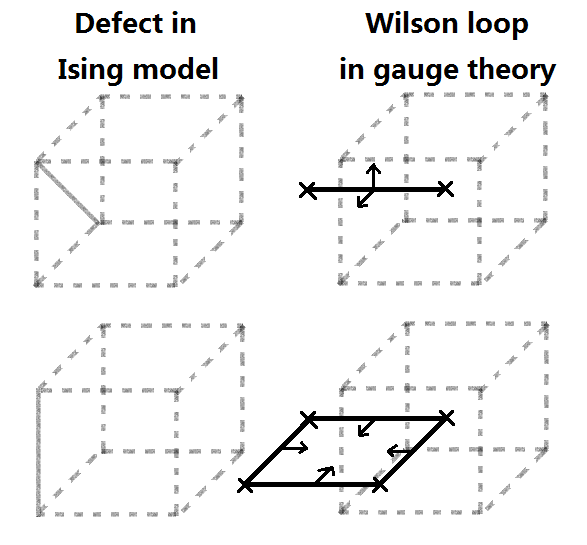}
  \caption{The duality between the defect bonds in the Ising model and the gauge invariant loops in the gauge theory. Note that a two-vector edge can also be seen as a loop of two single-vector edges, going back and forth.}
\end{figure}

We can also combine the exponentials in Eq.~13, and make Eq.~9 into
\begin{equation}
\frac{1}{Z_\Ising}\sum_\config \exp(-S_\Ising^*)=\frac{Z_\Ising^*}{Z_\Ising},
\end{equation}
where $S_\Ising^*=S_\Ising+2\beta_1 s_v^z s_{v'}^z$. Considering the definition of $S_\Ising$ in Eq.~1, we can interpret $S_\Ising^*$ as that the coupling coefficient between $v$ and $v'$ becomes $-\beta_1$, i.e. that there is a defect nn bond.

With similar methods, we can show that the average value of a two-vector edges in our gauge theory equals to the average value of $e^{-2\beta_2 s_v^z s_{v''}^z}$ where $v$ and $v''$ are nnn pair. It also equals to the partition function when there is a defect nnn bond.

In Fig.~8, we summarize the relation between defect bonds in the Ising model and the average value of Wilson loops in our gauge theory. Any more complicated Wilson loop is a combination of these two basic cases.

\section {Discussion}

In the previous sections, we have only talked about the 3D Ising model on an isotropic cubic lattice. Actually the duality even works for the very anisotropic case, in which there are three different nn coupling coefficients and six different nnn coupling coefficients. The dual gauge theory will have three different coefficients for the plaquettes in different orientations, and will have different coefficients for $[a_1 a_2]$ and $[a_2 a_3]$ in Eq.~\ref{S-gauge}. The dual relations keep simple as $\sinh(2\beta_{\cdots})\sinh(2\alpha_{\cdots})=1$.

An interesting anisotropic case is when nnn coupling only occurs in the x-y plane. The third dimension can be seen as the time direction: it becomes an Ising model with 2D Hamiltonian. Because the nnn coupling in the x-z and y-z planes is zero, the edge term coefficients in the y and x edges in the gauge theory become infinity, which force the four variables on those edges become the same. Therefore, the dual gauge theory only has vector variables on the z-direction edges. With this simplification, we can write down a Hamiltonian form for this gauge theory. We do not know whether there is a Hamiltonian form for the gauge theory we discussed in the previous sections.

The proof of duality in Sec.~\ref{Proof} can also be made with the method of path integral, however not as intuitive as the high temperature expansion and the low temperature expansion of the two models.

For the future work, we will use numerical simulations to explore how the phase transition point changes with the ratio between the nnn and nn couplings. We will investigate the scaling laws of the Wilson loops around the critical region, and try to find out whether there is universality in our model. Theoretical approach will also be helpful to understand the features of this model.

\begin {acknowledgments}
The author would like to express sincere gratitude to the mentor Prof. Anton Kapustin for the guidance. This work was supported by the Caltech SURF program. The author acknowledges the generous support of Dr. Neher and his family. Thanks to Yu-An Chen for many fruitful discussion. Some computations in this project were conducted on the Caltech High Performance Computing Cluster.
\end {acknowledgments}

\bibliography {library}

\begin{thebibliography}{9}%
\makeatletter
\providecommand \@ifxundefined [1]{%
 \@ifx{#1\undefined}
}%
\providecommand \@ifnum [1]{%
 \ifnum #1\expandafter \@firstoftwo
 \else \expandafter \@secondoftwo
 \fi
}%
\providecommand \@ifx [1]{%
 \ifx #1\expandafter \@firstoftwo
 \else \expandafter \@secondoftwo
 \fi
}%
\providecommand \natexlab [1]{#1}%
\providecommand \enquote  [1]{``#1''}%
\providecommand \bibnamefont  [1]{#1}%
\providecommand \bibfnamefont [1]{#1}%
\providecommand \citenamefont [1]{#1}%
\providecommand \href@noop [0]{\@secondoftwo}%
\providecommand \href [0]{\begingroup \@sanitize@url \@href}%
\providecommand \@href[1]{\@@startlink{#1}\@@href}%
\providecommand \@@href[1]{\endgroup#1\@@endlink}%
\providecommand \@sanitize@url [0]{\catcode `\\12\catcode `\$12\catcode
  `\&12\catcode `\#12\catcode `\^12\catcode `\_12\catcode `\%12\relax}%
\providecommand \@@startlink[1]{}%
\providecommand \@@endlink[0]{}%
\providecommand \url  [0]{\begingroup\@sanitize@url \@url }%
\providecommand \@url [1]{\endgroup\@href {#1}{\urlprefix }}%
\providecommand \urlprefix  [0]{URL }%
\providecommand \Eprint [0]{\href }%
\providecommand \doibase [0]{http://dx.doi.org/}%
\providecommand \selectlanguage [0]{\@gobble}%
\providecommand \bibinfo  [0]{\@secondoftwo}%
\providecommand \bibfield  [0]{\@secondoftwo}%
\providecommand \translation [1]{[#1]}%
\providecommand \BibitemOpen [0]{}%
\providecommand \bibitemStop [0]{}%
\providecommand \bibitemNoStop [0]{.\EOS\space}%
\providecommand \EOS [0]{\spacefactor3000\relax}%
\providecommand \BibitemShut  [1]{\csname bibitem#1\endcsname}%
\let\auto@bib@innerbib\@empty
\bibitem [{\citenamefont {Gaillard}\ \emph {et~al.}(1999)\citenamefont
  {Gaillard}, \citenamefont {Grannis},\ and\ \citenamefont
  {Sciulli}}]{Gaillard1999}%
  \BibitemOpen
  \bibfield  {author} {\bibinfo {author} {\bibfnamefont {M.~K.}\ \bibnamefont
  {Gaillard}}, \bibinfo {author} {\bibfnamefont {P.~D.}\ \bibnamefont
  {Grannis}},\ and\ \bibinfo {author} {\bibfnamefont {F.~J.}\ \bibnamefont
  {Sciulli}},\ }\href {\doibase 10.1103/revmodphys.71.s96} {\bibfield
  {journal} {\bibinfo  {journal} {Reviews of Modern Physics}\ }\textbf
  {\bibinfo {volume} {71}},\ \bibinfo {pages} {S96} (\bibinfo {year} {1999})},\
  \Eprint {http://arxiv.org/abs/hep-ph/9812285} {arXiv:hep-ph/9812285}
  \BibitemShut {NoStop}%
\bibitem [{\citenamefont {Wilson}(1974)}]{Wilson1974}%
  \BibitemOpen
  \bibfield  {author} {\bibinfo {author} {\bibfnamefont {K.~G.}\ \bibnamefont
  {Wilson}},\ }\href {\doibase 10.1103/physrevd.10.2445} {\bibfield  {journal}
  {\bibinfo  {journal} {Physical Review D}\ }\textbf {\bibinfo {volume} {10}},\
  \bibinfo {pages} {2445} (\bibinfo {year} {1974})}\BibitemShut {NoStop}%
\bibitem [{\citenamefont {Wegner}(1971)}]{Wegner1971}%
  \BibitemOpen
  \bibfield  {author} {\bibinfo {author} {\bibfnamefont {F.~J.}\ \bibnamefont
  {Wegner}},\ }\href {\doibase 10.1063/1.1665530} {\bibfield  {journal}
  {\bibinfo  {journal} {Journal of Mathematical Physics}\ }\textbf {\bibinfo
  {volume} {12}},\ \bibinfo {pages} {2259} (\bibinfo {year}
  {1971})}\BibitemShut {NoStop}%
\bibitem [{\citenamefont {Wegner}(2014)}]{Wegner2014}%
  \BibitemOpen
  \bibfield  {author} {\bibinfo {author} {\bibfnamefont {F.~J.}\ \bibnamefont
  {Wegner}}} (\bibinfo {year} {2014}),\ \Eprint
  {http://arxiv.org/abs/1411.5815} {arXiv:1411.5815} \BibitemShut {NoStop}%
\bibitem [{\citenamefont {Kramers}\ and\ \citenamefont
  {Wannier}(1941)}]{Kramers1941}%
  \BibitemOpen
  \bibfield  {author} {\bibinfo {author} {\bibfnamefont {H.~A.}\ \bibnamefont
  {Kramers}}\ and\ \bibinfo {author} {\bibfnamefont {G.~H.}\ \bibnamefont
  {Wannier}},\ }\href {\doibase 10.1103/physrev.60.252} {\bibfield  {journal}
  {\bibinfo  {journal} {Physical Review}\ }\textbf {\bibinfo {volume} {60}},\
  \bibinfo {pages} {252} (\bibinfo {year} {1941})}\BibitemShut {NoStop}%
\bibitem [{\citenamefont {Wannier}(1945)}]{Wannier1945}%
  \BibitemOpen
  \bibfield  {author} {\bibinfo {author} {\bibfnamefont {G.~H.}\ \bibnamefont
  {Wannier}},\ }\href {\doibase 10.1103/revmodphys.17.50} {\bibfield  {journal}
  {\bibinfo  {journal} {Reviews of Modern Physics}\ }\textbf {\bibinfo {volume}
  {17}},\ \bibinfo {pages} {50} (\bibinfo {year} {1945})}\BibitemShut {NoStop}%
\bibitem [{\citenamefont {Bill{\'{o}}}\ \emph {et~al.}(2013)\citenamefont
  {Bill{\'{o}}}, \citenamefont {Caselle}, \citenamefont {Gaiotto},
  \citenamefont {Gliozzi}, \citenamefont {Meineri},\ and\ \citenamefont
  {Pellegrini}}]{Bill2013}%
  \BibitemOpen
  \bibfield  {author} {\bibinfo {author} {\bibfnamefont {M.}~\bibnamefont
  {Bill{\'{o}}}}, \bibinfo {author} {\bibfnamefont {M.}~\bibnamefont
  {Caselle}}, \bibinfo {author} {\bibfnamefont {D.}~\bibnamefont {Gaiotto}},
  \bibinfo {author} {\bibfnamefont {F.}~\bibnamefont {Gliozzi}}, \bibinfo
  {author} {\bibfnamefont {M.}~\bibnamefont {Meineri}},\ and\ \bibinfo {author}
  {\bibfnamefont {R.}~\bibnamefont {Pellegrini}},\ }\href {\doibase
  10.1007/jhep07(2013)055} {\bibfield  {journal} {\bibinfo  {journal} {Journal
  of High Energy Physics}\ }\textbf {\bibinfo {volume} {2013}} (\bibinfo {year}
  {2013}),\ 10.1007/jhep07(2013)055},\ \Eprint {http://arxiv.org/abs/1304.4110}
  {arXiv:1304.4110} \BibitemShut {NoStop}%
\bibitem [{\citenamefont {Onsager}(1944)}]{Onsager1944}%
  \BibitemOpen
  \bibfield  {author} {\bibinfo {author} {\bibfnamefont {L.}~\bibnamefont
  {Onsager}},\ }\href {\doibase 10.1103/physrev.65.117} {\bibfield  {journal}
  {\bibinfo  {journal} {Physical Review}\ }\textbf {\bibinfo {volume} {65}},\
  \bibinfo {pages} {117} (\bibinfo {year} {1944})}\BibitemShut {NoStop}%
\bibitem [{\citenamefont {Pelissetto}\ and\ \citenamefont
  {Vicari}(2002)}]{Pelissetto2002}%
  \BibitemOpen
  \bibfield  {author} {\bibinfo {author} {\bibfnamefont {A.}~\bibnamefont
  {Pelissetto}}\ and\ \bibinfo {author} {\bibfnamefont {E.}~\bibnamefont
  {Vicari}},\ }\href {\doibase 10.1016/s0370-1573(02)00219-3} {\bibfield
  {journal} {\bibinfo  {journal} {Physics Reports}\ }\textbf {\bibinfo {volume}
  {368}},\ \bibinfo {pages} {549} (\bibinfo {year} {2002})},\ \Eprint
  {http://arxiv.org/abs/cond-mat/0012164} {arXiv:cond-mat/0012164} \BibitemShut
  {NoStop}%
\end{thebibliography}%

\end {document}